\documentclass[aps,prl,superscriptaddress,longbibliography,twocolumn,floatfix]{revtex4-2}
\usepackage{CJK}
\usepackage[table,cmyk]{xcolor}
\usepackage{amsmath, amsthm,commath,braket,booktabs,dsfont}
\usepackage{calc,graphicx,bm,tikz}
\usepackage[charter,cal=cmcal,sfscaled=false]{mathdesign}
\usepackage[percent]{overpic}
\usepackage{xr}

\usepackage{multirow}
\definecolor{red}{cmyk}{0.12,0.94,0.87,0.34}
\definecolor{blue}{cmyk}{0.99,0.78,0.16,0.03}

\usepackage{hyperref}

\hypersetup{
  pdftitle={Harnessing Non-convex Quantum Correlations of Independent Qubits},
  pdfauthor={Sun, Zhou, Zhang et al.},
  pdfstartview=Fit,
  pdfpagelayout=SinglePage,
  colorlinks,
  linkcolor=blue,
  citecolor=blue,
  urlcolor=red}

\newtheorem{theorem}{Theorem}
\newtheorem{proposition}[theorem]{Proposition} 
\begin{document}
\begin{CJK*}{UTF8}{}

\title{Harnessing Non-convex Quantum Correlations of Independent Qubits}

\author{Liang-Liang Sun (\CJKfamily{gbsn}孙亮亮)}
%\email{sun18@ustc.edu.cn}
\affiliation{Department of Modern Physics and National Laboratory for Physical Sciences at Microscale, University of Science and Technology of China, Hefei, Anhui 230026, China}

\author{Xiang Zhou}
\affiliation{Department of Modern Physics and National Laboratory for Physical Sciences at Microscale, University of Science and Technology of China, Hefei, Anhui 230026, China}

\author{Chengjie Zhang}
\affiliation{School of Physical Science and Technology, Ningbo University, Ningbo, 315211, China}

\author{Zizhu Wang (\CJKfamily{gbsn}王子竹)}
\email{zizhu@uestc.edu.cn}
\affiliation{Institute of Fundamental and Frontier Sciences, University of Electronic Science and Technology of China, 611731 Chengdu, China.}
\affiliation{Key Laboratory of Quantum Physics and Photonic Quantum Information (University of Electronic Science and Technology of China), Ministry of Education, 611731, Chengdu, China}

\author{Yong-Shun Song}
\email{ys.song320@gmail.com}
\affiliation{School of Information Engineering, Changzhou Vocational Institute of Industry Technology, Changzhou 213164, China.}

\author{Sixia Yu (\CJKfamily{gbsn}郁司夏)}
\email{yusixia@ustc.edu.cn} 
\affiliation{Department of Modern Physics and National Laboratory for Physical Sciences at Microscale, University of Science and Technology of China, Hefei, Anhui 230026, China}
\affiliation{Hefei National Laboratory, University of Science and Technology of China, Hefei 230088, China}

\begin{abstract}
Quantum correlations in Bell and prepare-and-measure experiments are central resources for probing nonclassicality and enabling device-based quantum information protocols. In the absence of shared public randomness (i.e., without run-to-run mixing), even qubit correlation sets are typically non-convex, making standard convex characterizations inadequate. Here we derive qubit-specific constraints from uncertainty relations, yielding a state-independent consistency test for observed statistics in both prepare-and-measure and Bell scenarios. The test captures explicit non-convex boundaries in representative correlation families and enables correlation-based device inference by constraining (and sometimes uniquely determining) unitary-invariant measurement parameters even away from extreme points. Moreover, incorporating the inferred qubit constraints as additional conditions in a moment-matrix relaxation strengthens separability tests and can certify entanglement even for Bell-local correlations within the independent-device model. These tools provide a practical route to characterize and leverage low-dimensional quantum devices, including certification, randomness generation, and entanglement verification.
\end{abstract}

\maketitle
\end{CJK*}

\noindent\textit{Introduction.---}Quantum correlations observed in Bell tests and prepare-and-measure (PM) experiments are a cornerstone of quantum foundations and a practical resource for quantum information processing~\cite{1975AmJPh, RevModPhys.82.665, PhysRevLett.67.661, Joseph2022, Acc2006, scienceabe8770, PhysRevLett.68.557, RevModPhys.86.419, Buhrman2010},  enabling tasks such as device-independent cryptography and randomness generation beyond classical capabilities~\cite{RevModPhys.86.419, PhysRevLett.97.120405, PhysRevLett.114.150501}.
In many near-term platforms, the relevant systems are effectively qubits~\cite{Pelucchi2022,Pla2013, RevModPhys.76.1037, do1372, Kloeffel2013, traped, Saffman2016, Zhou2014, Doherty2013, Sarma2015, Pelucchi2022, Pla2013} , and the structure of achievable correlations is directly tied to the operational performance and certification of quantum technologies~\cite{PhysRevLett.97.120405, PhysRevLett.115.150501, PhysRevLett.88.210401, do1372, PhysRevLett.95.080501, PhysRevA.92.042117,PhysRevA.90.022322, PhysRevA.100.030301,PhysRevLett.117.260401, Tavakoli2020, PhysRevResearch.2.033014, PhysRevLett.121.050501,Tavakoli2024} . Therefore, understanding this structure is an essential issue in quantum information science.

% in diverse information-processing tasks over classical counterparts~\cite{1975AmJPh, RevModPhys.82.665, PhysRevLett.67.661, Joseph2022, Acc2006, scienceabe8770, PhysRevLett.68.557, RevModPhys.86.419, Buhrman2010}.

The core challenge of this issue stems from the typically non-convex nature of the qubit correlation space~\cite{PhysRevA.80.042114, PhysRevA.92.062120, Lanyon2009, PhysRevLett.123.070505, PhysRevLett.113.230501}, which necessitates a nonlinear analytical approach. Current research  often  adapts existing linear approaches---such as semidefinite programming and moment-matrix hierarchies (e.g. the Navascu\'{e}s-Pironio-Ac\'{i}n (NPA)  hierarchy~\cite{PhysRevLett.98.010401,NPA2} and its variants~\cite{PhysRevLett.100.210503, PhysRevX.4.011011, PhysRevLett.115.020501, PhysRevLett.129.250504}) to this problem via assuming  the shared public randomness between devices, which  allows for the run-to-run mixing of correlations from different implementations, thereby convexifying the correlation set. These linear approaches effectively demonstrate how dimensionality imposes constraints on both Bell's nonlocality ~\cite{PhysRevLett.119.080401, PhysRevA.78.062112}  and information-theoretic capabilities in specific tasks, such as secure communication~\cite{PhysRevLett.97.120405, PhysRevLett.115.150501} and quantum random access codes~\cite{PhysRevLett.88.210401, do1372, PhysRevLett.95.080501}. However, they inherently target convex relaxations, where the permitted mixing process may obscure quantum features rooted in the underlying non-convex structure.  For instance, certain local correlations may also require entangled states~\cite{Le2023quantumcorrelations}. Despite this relevance, efficient tools for the independent-device (non-convex) regime remain limited, leaving two questions open: (i) how to characterize correlations achievable by independent qubit devices in Bell and PM scenarios, and (ii) how to leverage such characterizations for device inference and entanglement detection without requiring extreme correlations or Bell nonlocality.

Uncertainty relations provide a promising approach to study qubit correlations. Historically, nonlocal correlations emerged from the challenge to the indeterminism of the uncertainty principle~\cite{PhysRev.47.777, bell}. The recognition that quantum nonlocality does not reach the maximum no-signaling bounds~\cite{Popescu1994} has spurred the search for explanatory physical principles~\cite{vanDam2005} (see \cite{RevModPhys.86.419} for a review), with uncertainty relations regarded as a key candidate~\cite{Oppenheim2010}. This principle has been used to explain quantum bounds in minimal Bell scenarios~\cite{PhysRevA.106.032213, PhysRevA.109.022408, hsu}, but it falls short of addressing the core issue: how to characterize quantum correlations arising from independent qubit systems.

In this Letter we answer the two core questions by exploiting the geometry of qubit measurements encoded in the uncertainty relation~\cite{li2015r}. Specifically, we first derive a qubit uncertainty-relation-based correlation criterion that enforces a state-independent consistency constraint on observed expectation values. We then use it to obtain explicit non-convex boundaries in examples and compare with representative SDP and nonlinear approaches~\cite{PhysRevLett.98.010401, PhysRevLett.112.140407, PhysRevLett.117.060401}. We introduce a \emph{correlation-based device inference} (CBDI) protocol that bounds (and in certain regimes uniquely determines) unitary-invariant measurement parameters  and finally strengthen entanglement detection by incorporating the inferred parameters into a moment-matrix test, enabling entanglement certification even for Bell-local correlations within the independent-device model~\cite{PhysRevLett.98.010401, Le2023quantumcorrelations}. Our approach provides a practical tool for analyzing correlations in independent qubit systems and harnessing them for device inference and entanglement detection.

  %is a  complementary to SDP/NPA-based tools~\cite{PhysRevLett.98.010401, PhysRevLett.115.020501} and to existing nonlinear, dimension-bounded witnesses based on determinants or purity~\cite{PhysRevLett.112.140407, PhysRevLett.117.060401}.  

\medskip
\noindent\textit{Problem setup.---} Quantum correlations arising from independent devices are those generated by performing measurements \(\{\{\mathbf{M}_{\mathbf{a}|\mathbf{s}}\}_{\bf a, \bf s}\) on a quantum state \(\rho\):
\[
\mathbf{P} = \bigl\{ p(\mathbf{a}|\mathbf{s}) = \operatorname{Tr}\bigl(\rho \,\mathbf{M}_{\mathbf{a}|\mathbf{s}}\bigr) \bigr\}_{\mathbf{a},\mathbf{s}},
\]
where run-to-run mixing is forbidden and \(\mathbf{s}\) (\(\mathbf{a}\)) denote the measurement settings (outcomes). By contrast, if the devices share public randomness $\lambda$, one may mix implementations,
$\bigl\{ p(\mathbf{a}|\mathbf{s}) = \sum_\lambda \operatorname{Tr}\bigl(\rho_\lambda \mathbf{M}_{\mathbf{a}|\mathbf{s},\lambda}\bigr) \, p(\lambda) \bigr\}_{\mathbf{a},\mathbf{s}},$
which convexifies the correlation set. For finite-dimensional systems these two models differ, and the independent-device set is typically non-convex.

We focus on correlations arising from independent qubit systems, beginning with  a PM scenario. It consists of a source and a measurement device. The source can select a qubit state $\rho_{k}$ from a set $\Omega:= \{\rho_{k}\}$  to  prepare, and then transmit it to the measurement device. The measurement device then selects a  setting $A_{i}$ from a set $\mathcal{O}:= \{A_{i}\}$ to perform, yielding an outcome $a$ with probability denoted  by  $p(a|A_i, \rho_{k})$. To introduce the correlation criterion, we begin with binary projective measurements (PVMs) of the form $A_i=\vec a_i\cdot\vec\sigma$, where $\vec{a}_{i}$ is a unit vector specifying the measurement direction and $\vec{\sigma}$ denotes the Pauli vector.  For two PVMs  $A_{i}$ and $A_{j}$,   uncertainty relation~\cite{li2015r} constrains their expectation values $A_{i|\rho}:= {\rm Tr}(\rho A_{i})$  as
\begin{align} % 使用 align 环境
A_{i|\rho}^2 +A_{j|\rho}^2 + c_{ij}^2 - 2c_{ij} A_{i|\rho} A_{j|\rho} \leq 1,\label{equr}
\end{align}
where $A^{2}_{i|\rho}=[{\rm Tr}(\rho A_{i})]^{2}$ and   $c_{ij} := \vec{a}_i \cdot \vec{a}_j$, referred to as measurement overlap here,   is the cosine of the relative angle between the $\vec{a}_{i}$ and $\vec{a}_{j}$,  which   characterizes these two PVMs up to a unitary.   By reformulating Eq.(\ref{equr}) into $(c_{ij}-A_{i|\rho} A_{j|\rho})^{2}\leq (1-A_{i|\rho}^{2})(1- A_{j|\rho}^{2})$, it is easy to see that   $ g_{- }( A_{i|\rho}, A_{j|\rho})\leq  c_{ij}  \leq  g_{+ }(A_{i|\rho}, A_{j|\rho})$ with    $g_{\pm }( A_{i|\rho}, A_{j|\rho}):= {\Pi}_{l=i, j}A_{l|\rho} \pm {\Pi}_{l=i, j}\sqrt{1- A_{l|\rho}^{2}}$. Henceforth, we will omit $A_{i|\rho}$ and  $A_{j|\rho}$ in the function $g_{\pm}$ for brevity unless necessary. Since $c_{ij}$ is state independent, the intervals obtained from different $\rho_k\in\Omega$ must have a nonempty intersection, which yields the following correlation criterion.

\begin{theorem}[Uncertainty-relation (UR) criterion for PVM]
 If  expectation values   \(\{A_{i|\rho_{k}} \mid A_{i}\in \mathcal{O},  \rho_{k}\in\Omega\}_{a}\) arise from a PM scenario  using  binary qubit PVMs \(A_{i}\) and \(A_{j}\) in the absence of public randomness, then one has
\begin{eqnarray}
\max_{\rho_{k}\in \Omega}g_{-} (A_{i|\rho_{k}}, A_{j|\rho_{k}})
\leq\min_{\rho_{k}\in \Omega}g_{+}( A_{i|\rho_{k}}, A_{j|\rho_{k}}). \label{g3a_1}
\end{eqnarray}
\end{theorem}
Because $\{ A_{i|\rho_k}\}_{i,k}$ are directly computed from experimental frequencies, this theorem can be applied immediately to PM data.

We now consider general binary qubit measurements described by positive operator-valued measures (POVMs) as $A_i = r_i' \, \vec{a}_i \cdot \vec{\sigma} + r_i \mathbb{I},$ where the positivity of the measurement elements \( (\mathbb{I} \pm A_i)/2 \) requires $0 \le r_i' \le 1, \quad -1 \le r_i \le 1, \quad \text{and} \quad r_i' + |r_i| \le 1.$ Thus, two POVMs $A_{i}$  and  $A_{j}$  are specified with  parameters \(\theta_{5} := \{ r_i, r_j, r'_i, r'_j, c_{ij} \}\)  up to a unitary. For later use, we also define  notions \(\theta_{4} := \{ r_i, r_j, r'_i, r'_j \}\) and \(\tilde\theta_{4} := \{ r_i, r_j, \bar r_i, \bar r_j \}\) with \(\bar r_l := 1 - |r_l|\).  Note that  \((A_i - r_i \mathbb{I})/r'_i\) and \((A_j - r_j \mathbb{I})/r'_j\) are themselves PVMs,  substituting them into the uncertainty relation Eq. (\ref{equr}) leads to  $g^{(\theta_{4})}_- \le  c_{ij} r'_i r'_j  \le  g^{(\theta_{4})}_+,$ where $g^{(\theta_{4})}_\pm 
:= \prod_{l=i,j} \bigl( A_{l|\rho} - r_l \bigr) \pm \prod_{l=i,j} \sqrt{ r^{\prime 2}_l - \bigl( A_{l|\rho}- r_l \bigr)^2 }.$
Since \( r'_l \le  \bar r_l\), we have $g^{(\tilde \theta_{4})}_- \le g^{(\theta_{4})}_- \quad \text{and} \quad g^{(\theta_{4})}_+ \le g^{(\tilde \theta_{4})}_+$.  Using $r'_l\le \bar r_l:=1-|r_l|$ yields the relaxed bounds
$g_-^{(\tilde\theta_4)}\le g_-^{(\theta_4)}$ and $g_+^{(\theta_4)}\le g_+^{(\tilde\theta_4)}$, which eliminate the explicit dependence on $r'_i,r'_j$ and give a state-independent necessary condition. This relaxation may be conservative (as  the relevant measurement $r_l' \, \vec{a}_l \cdot \vec{\sigma} + \bar r_l \mathbb{I}$ is well-defined ), but it is sufficient for the correlation tests and inference tasks considered here (see appendix for details). This interval constraint on $c_{ij}r'_{i}r'_{j}$ holds for all $\rho_{k} \in \Omega$,  implying  a correlation criterion  formulated in terms of \(\{ r_i, r_j\}\):

\begin{theorem}[Uncertainty relation (UR)-based correlation criterion]
If expectation values   \(\{ A_{i|\rho_{k}} \mid A_{i}\in \mathcal{O},  \rho_{k}\in\Omega\}_{i, k}\) arise from a PM scenario using  binary qubit POVMs \(A_{i}\) and \(A_{j}\) in the absence of public randomness, then there  exist two real numbers \(-1\le r_{i}, r_{j} \le 1\) such that
\begin{align} % 使用 align 环境
\max_{\rho_{k}\in \Omega} g^{(\tilde \theta_{4})}_{-} (A_{i|\rho_{k}}, A_{j|\rho_{k}})
&\leq\min_{\rho_{k}\in \Omega}g^{(\tilde \theta_{4})}_{+} (A_{i|\rho_{k}}, A_{j|\rho_{k}}),  \label{g3a_2}
\end{align}
where $\tilde{\theta}_4 = \{r_i, r_j, \bar{r}_i, \bar{r}_j\}$ with $\bar{r}_\ell = 1 - |r_\ell|$ is the set of measurement parameters.
\end{theorem}
Operationally, substituting $A_{i|\rho_{k}}$ and $A_{j|\rho_{k}}$ into $g^{(\tilde\theta_{4})}_{\pm}$, one has an inequality of  $r_i$ and $r_j$, deciding whether solutions exist reduces to a low-dimensional feasibility problem and can be done in time polynomial in $|\Omega|$ and the input precision (see~appendix  for more details).  This criterion can be readily generalized to multiple measurements and to multi-outcome POVMs: for a measurement with $d$ outcomes $\{M_i\}_{i=1}^{d}$, one can decompose it into $d$ binary measurements of the form $\{M_i, \mathbb{1} - M_i\}$.  For a set of binary measurements, one can apply  the criterion to all possible pairs of them.

\medskip
\noindent\textit{ Criterion for qubit correlations in the Bell scenario.---}In a Bell scenario, $n$ observers share an $n$-partite state $\rho$. Party $m$ chooses a setting $s_m\in \mathcal{O}_m$ and obtains an outcome $a_m$, described by measurement operators $M^{a_m}_{s_m}$.
 Denoting the collections of settings and outcomes by $\mathbf{s} := (s_1,\dots,s_n)$ and $\mathbf{a} := (a_1,\dots,a_n)$,  the joint probability distribution is given by $ \{p(\mathbf{a}|\mathbf{s})= \operatorname{Tr}\bigl(\rho \bigotimes_{m=1}^n M^{a_m}_{s_m}\bigr)\}_{\mathbf{a}, \bf{s}}$. 

To bound correlations when party $m$ holds a qubit, we reduce the Bell data to a PM instance via conditional states: depending on the measurement settings $\mathbf{s}_{\bar m} := (\dots, s_l, \dots )_{l\neq m}$ and the outcomes $\mathbf{a}_{\bar m} := (\dots, a_l, \dots )_{l\neq m}$ of the other $n-1$ parties, a conditional state
\[
\rho_{\mathbf{a}_{\bar m}|\mathbf{s}_{\bar m}} = \frac{\operatorname{Tr}_{\bar m}\bigl[\rho \bigotimes_{l\neq m} M^{a_l}_{s_l}\bigr]}{p(\mathbf{a}_{\bar m}|\mathbf{s}_{\bar m})}
\]
is prepared for the $m$-th party with probability $p(\mathbf{a}_{\bar m}|\mathbf{s}_{\bar m}) = \sum_{a_m} p(a_m, \mathbf{a}_{\bar m} \mid s_m, \mathbf{s}_{\bar m})$. On this local state, the expectation value of a  binary measurement $A_i\in \mathcal{O}_{m}$ reads 
$ A_{i| \rho_{\mathbf{a}_{\bar m}|\mathbf{s}_{\bar m}}} = \frac{p(0, \mathbf{a}_{\bar m} \mid i, \mathbf{s}_{\bar m}) - p(1, \mathbf{a}_{\bar m} \mid i, \mathbf{s}_{\bar m})}{p(\mathbf{a}_{\bar m} \mid \mathbf{s}_{\bar m})}$, which is subject to the PM correlation criterion where   $\Omega_{m}= \{\rho_{\mathbf{a}_{\bar m} \mid \mathbf{s}_{\bar m}}\}_{{\mathbf{a}_{\bar m} \mid \mathbf{s}_{\bar m}}}$ and  $\mathcal{O}_{m}=\{A_{s_{m}}\}$. We next compare the resulting uncertainty relation-based criteria with representative linear (SDP) and nonlinear benchmarks in both Bell and PM settings.

 \begin{figure}\centering 
\includegraphics[scale=0.705]{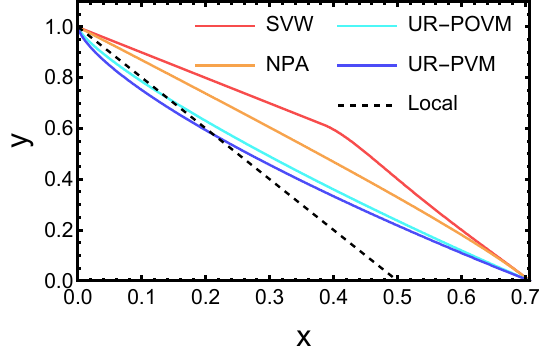}\\
\caption{Characterizing correlations   $Q^{Bell}_{xy}$  featuring independent  qubit systems  with  Sikora-Varvitsiotis-Wei (SVW) witness,    Navascu\'{e}s-Pironio-Ac\'{i}n (NPA) criterion, and uncertainty relation  (UR-PVM and UR-POVM),  which  exclude the region above curves.  Local correlation is also plotted (dashed line).}
\end{figure}

We first consider the minimal Bell scenario, where there are  two parties, each with two dichotomic measurements $A_\alpha$ and $B_\beta$ ($\alpha,\beta \in \{0,1\}$) to choose from, with outcomes $a,b \in \{0,1\}$ . We examine the following family of correlations parameterized by $(x,y)$:
\begin{equation}
Q^{\text{Bell}}_{xy} = x \mathcal{P}_{\text{PR}} + y \mathcal{P}_{\text{L}} + (1-x-y)\mathcal{P}_{\text{w}},
\end{equation}
which consists of three components: (1)  the extreme non-signaling correlation~\cite{Popescu1994}  $\mathcal{P}_{\text{PR}} = \{p(ab|\alpha\beta) = \frac{1}{2} \text{ if } a \oplus b = \alpha \beta\}$, (2) a local extreme correlation $\mathcal{P}_{\text{L}} = \{p(00|\alpha\beta) = 1, \forall \alpha, \beta\}$, and (3) the white noise  $\mathcal{P}_{\text{w}} = \{p(ab|\alpha\beta) = \frac{1}{4}, \forall a, b, \alpha, \beta\}$. This family of correlations can be realized with independent qubit systems only for some special pairs $\{x, y\}$. As  illustrated in Fig.~1,  our criteria, the nonlinear  Sikora-Varvitsiotis-Wei  (SVW) approach~\cite{PhysRevLett.117.060401}, and  the linear NPA hierarchy~\cite{PhysRevLett.98.010401,NPA2} give distinct forbidden zones (see~appendix  for more details), which lie above their respective curves. Our criteria  successfully capture the non-convex structure of quantum correlations. Notably, part of the Bell-local region (dashed line) lies outside the independent-qubit set and therefore cannot be generated by independent qubit devices.

 \begin{figure}\centering 
\includegraphics[scale=0.69]{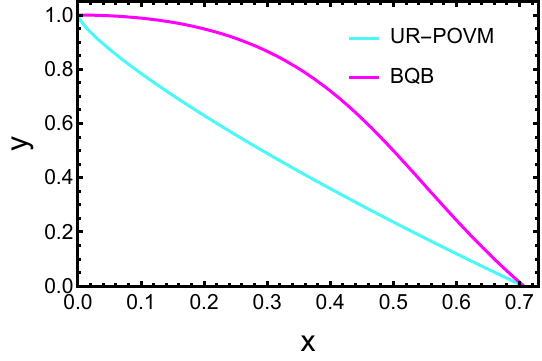}\\
\caption{Characterizing correlations $Q^{PM}_{xy}$ featuring independent  qubit systems with  the Bowles-Quintino-Brunner (BQB)  witness and uncertainty relation considering POVMs,  where the regions above these curves are forbidden by these criteria. }
\end{figure}

In the PM setting, we compare our UR criterion with the determinant-based Bowles--Quintino--Brunner (BQB) witness~\cite{PhysRevLett.112.140407} on the family $Q^{\mathrm{PM}}_{xy}$ (defined in appendix ), obtained from the above Bell family by identifying Alice's conditional states $\rho_{b|\beta}$. Figure~2 shows the resulting boundary; for this family the UR-POVM boundary is tight in the sense that boundary points admit explicit qubit realizations (see~appendix  for more details). While the nonlinear QBQ and SVW criteria are not sufficiently tight to produce nonconvex boundaries in examples above, they remain applicable to systems of any dimension to verify dimensionality. Our criterion complements these approaches by tailoring to qubit systems and enabling the exploration of nonconvex boundaries.

%In  PM scenario, we compare our criterion with the determinant-based criterion given  by Bowles-Quintino-Brunner (BQB) \cite{PhysRevLett.112.140407}. For this purpose, we consider $Q^{PM}_{xy}$ (see SM), which is a modification of  the above Bell scenario, with $\mathcal{O} = \{A_{0}, A_{1}\}$ and $\Omega = \{\rho_{b|\beta}\}$, where $\rho_{b\mid\beta}$ denotes the state on Alice's side conditioned on Bob's measurement $B_{\beta}$ and outcome $b$. In Fig. 2, we plot the determined $x-y$ boundary, where those relevant to uncertainty relations are optimal in the sense that the underlying states and measurements can be explicitly constructed (see SM).

\medskip
\noindent\textit{Applications: correlation-based device inference.---}A central task in quantum information science is to characterize devices from limited assumptions. Standard tomography methods require completely trusted states to characterize measurements, and trusted measurements to characterize states. Correlation-based certification relaxes these requirements by exploiting correlations that single out a unique implementation in ideal conditions (self-testing). However, self-testing typically relies on near-extremal correlations and yields essentially a yes/no verdict about a target configuration; if the test fails, it often provides limited diagnostic information.

In independent-device scenarios, run-to-run mixing is not permitted, which can dramatically reduce the set of device models compatible with a given correlation. This motivates a complementary approach, which we call correlation-based device inference (CBDI): given observed statistics, we infer feasible regions-and sometimes a unique value-of unitary-invariant measurement parameters, and then reconstruct compatible states.

Measurement  parameters $\theta_{5}$ relevant to $A_{i}$ and $A_{j}$ can be determined  in two steps: 
\begin{itemize}
    \item \textbf{Determine $r_{i}$ and $r_{j}$}: Substitute the expectation values $\{A_{i|\rho_{k}}\}_{i, k}$ into $g^{(\tilde{\theta}_{4})}_{\pm}$. Solving the inequality $\max_{\rho_{k} \in \Omega} g^{(\tilde{\theta}_{4})}_{-} \leq \min_{\rho_{k} \in \Omega} g^{(\tilde{\theta}_{4})}_{+}$ then gives the possible ranges for $r_{i}$ and $r_{j}$.
    \item \textbf{Determine $r'_{i}$, $r'_{j}$, and $c_{ij}$}: Substitute the possible  $r_{i}$ and $r_{j}$ from the previous step, along with $\{A_{i|\rho_{k}}\}_{i, k}$, into $g^{(\theta_{4})}_{\pm}$. Solving the inequality $\max_{\rho_{k} \in \Omega} g^{({\theta}_{4})}_{-} \leq \min_{\rho_{k} \in \Omega} g^{( {\theta}_{4})}_{+}$ yields the range of $r'_{i}$ and $r'_{j}$. Further solving $\max _{\rho_{k} \in \Omega} g^{(\tilde {\theta}_{4})}_{-} \leq c_{ij}r_{i}'r_{j}'\leq \min_{\rho_{k} \in \Omega}  g^{(\tilde {\theta}_{4})}_+$ gives $c_{ij}$.
\end{itemize}

Determining measured states is straightforward.  
%Given $\theta_{5}$ and the measurement set $\{A_{i|\rho_{k}} \}$, the possible quantum states can be inferred. 
One can  consider PVMs without loss of generality, as a binary  POVM can be transformed into a PVM. Specifically, the expectation values of two PVMs, $A_{i|\rho}$ and $A_{j|\rho}$, directly determine the Bloch vector $\vec s$ of the measured qubit state  \(\rho = \frac{I + \vec{s} \cdot \vec{\sigma}}{2}\) .
\begin{eqnarray}
\vec s&=&\frac{(A_{i|\rho}\vec a_j -A_{j|\rho}\vec a_i)\times(\vec a_i\times\vec a_j)+t(\vec a_i\times\vec a_j)}{(\vec a_i\times\vec a_j)^2}\nonumber
\end{eqnarray}
 where one can let $\vec{a}_{i}=(1, 0, 0)$ and $\vec{a}_{j}=(c_{ij}, \sqrt{1-c^{2}_{ij}}, 0)$, and  the additional parameter  satisfying $|t|\le \sqrt{(1-  A_{i|\rho}^2)(1- A_{j|\rho}^2)-(\vec a_i\cdot\vec a_j-A_{i|\rho} A_{j|\rho})^2}$  can be certified with   another measurement incompatible with $A_{i}$ and $A_{j}$.  As an example, we consider correlations $Q^{PM}_{xy}$ on the $x-y$ boundary in~appendix . From the correlations, we can uniquely determine two measurements $A_{i}$ and $A_{j}$  and three of the four states $\{\rho_{b|\beta}\}$, while the last state is determined up to  one free parameter. %In Bell's scenario, our criterion is  not tight as it does not incorporate the additional  requirement needed  to reconstruct a global state with local conditional states; a typical one is the non-signaling condition, namely,  $\sum_{\mathbf{a}_{\bar m}}  p(\mathbf{a}_{\bar  m}|\mathbf{s}_{\bar  m})\rho_{\mathbf{a}_{\bar  m}|\mathbf{s}_{\bar  m}}$ is independent of ${\bf {s}}_{\bar m}$. 

At first sight, one might attempt a brute-force approach by parameterizing all states and measurements and solving for consistency with the observed statistics. This leads to multivariate polynomial constraints that are NP-hard in general and quickly become impractical as the number of settings and states grows. In contrast, CBDI decomposes the inference into low-dimensional steps: for each measurement pair it reduces to feasibility of inequalities in a constant number of real parameters, yielding an efficient procedure (see~appendix  for more details).

Now, we compare CBDI with conventional linear self-testing protocols.  First, linear methods rely on extreme correlations specific to unique devices, yielding only yes-or-no answers. In contrast, CBDI applies to general correlations and can refine device parameters into feasible regions  using uncertainty relations. Second,   since uncertainty relations directly involve measurement parameters and experimental statistics, CBDI can refer  devices with  simple experimental setups. For instance, in the PM scenario, generating extreme correlations requires at least four states and three measurements~\cite{Drotos2024}. However, CBDI can uniquely determine  measurements and states with simpler devices.  %with only two states and two measurements, where two states are chosen such that that they saturate the upper and lower bounds of Eq.~(\ref{g3a_2}). 
Such examples are easy to find. For example, consider the correlation \( Q^{PM}_{xy} \) on the \( x-y \) boundary and the expectation values of $ \mathcal{O}=\{ A_0, A_{1}\}$ with respect to   $\Omega=\{\rho_{0|0}, \rho_{1|0}, \rho_{1|1}\}$. These relevant statistics  in this context  are sufficient to uniquely determine the measurements and states. Furthermore, to verify non-projective measurements, the linear approach relies on witnesses for which non-projective measurements yield larger values than PVMs---only a few such cases have been identified in Refs~\cite{PhysRevLett.117.260401, Tavakoli2020, Smania20}. In contrast, CBDI certifies nonprojectivity whenever the inferred $r'_{i}\neq 1$.

\medskip
\noindent\textit{Optimization of entanglement detection---}Detecting entanglement from correlation data is a central task in quantum information science. Bell-inequality violation certifies entanglement, but many entangled states produce Bell-local correlations. When partial information about the measurements is available, one can tighten separable-state bounds beyond the Bell-local bound~\cite{PhysRevA.85.032301, msm46n3}. Here we strengthen entanglement detection by incorporating CBDI-inferred qubit measurement relations as additional linear constraints in a moment-matrix (NPA-type) test. %Detecting entanglement is a central task in quantum information science. A standard tool for this purpose is Bell's test, where a violation of Bell's inequality verifies the presence of entanglement. When the employed measurement  are known, the bound for separable states can be tightened beyond the local bound.   For example, with $A_{0}=\sigma_{x}$ and $A_{1}=\sigma_{y}$, the bound of CHSH inequality  for separable states can be tightened to $\sqrt{2}$ from classical bound of $2$~\cite{PhysRevA.85.032301, msm46n3}. In this section, we  optimize the entanglement detection by incorporating the measurement information inferred in the previous section into the NPA hierarchy.

The NPA hierarchy~\cite{PhysRevLett.98.010401,NPA2} characterizes quantum correlations via a moment (Gram) matrix $\Gamma$ with entries
$\Gamma_{ef}:=\mathrm{Tr}(\rho\,O_e^\dagger O_f)$ for an operator list $\tilde O=\{O_e\}$ (e.g., products of measurement operators).
The observed correlators fix a subset of entries (``physical'' terms), while the remaining entries (``non-physical'' terms) are free variables.
A correlation is compatible with quantum theory if there exists an assignment to the non-physical terms such that $\Gamma\succeq 0$ (equivalently, its real part $\Gamma'=(\Gamma+\Gamma^*)/2\succeq 0$).

%The NPA hierarchy \cite{PhysRevLett.98.010401} is  a sequence of tests designed  to characterize correlations that can be realized with  a quantum system. It considers a set $\tilde O=\{O_{e}\}$ of operators $O_{e}$, e.g.,  some products of the measurement operators $\{M^{a_{i}}_{s_{i}}\}$ or linear combinations of them. Then their Gram matrix $\Gamma$ defined with  the entries  $\Gamma_{ef}:= {\rm Tr}(\rho O^{\dagger}_{e} O_{f})$ is positive semidefinite. Entries  relevant to observable statistics and thus having operational meaning are referred to as  physical terms, while the remaining ones involve non-commuting quantities and are called non-physical terms.  One commonly considers a real moment matrix, $\Gamma'=(\Gamma+\Gamma^{*})/2$, where the asterisk specifies  the complex conjugate. If a correlation can be generated with quantum systems, one can assign some real values on non-physical terms  such that $\Gamma'$ is semidefinite.   

Here, we consider that the $m$-th party holds a two-dimensional quantum system and,  relabel the elements of $\tilde O$ according to this partition as $O_{e,f}=O^{(m)}_{e}\otimes O^{(\bar m)}_{f}$, where $O^{(m)}_{e}$ involves measurement operators on  the $m$-th party while $O^{(\bar m)}_{f}$ for remaining parties. Then, we  define a matrix with the entry relevant to $O_{ef}$ and $O_{e'f' }$ as
\begin{equation}
\label{eq:gamma_def_split} % 如果断行，建议更新标签
\begin{split}
4\Gamma_{ee', ff'}=&{\rm Tr}\left[\rho \left(O^{(m)\dagger}_{e} {O}^{(m)}_{e'}+O^{(m)\dagger}_{e'} {O}^{(m)}_{e} \right)\right. \\
&\quad \left. {}\times \left(O^{(\bar m)\dagger}_{f} {O}^{(\bar m)}_{f'}+O^{(\bar m)\dagger}_{f'} {O}^{(\bar m)}_{f}\right)\right].
\end{split}
\end{equation}
For product states the moment matrix factorizes across the bipartition, and for separable states it lies in the convex hull of such factorizable matrices; in both cases the corresponding constraints imply positive semidefiniteness.
Entangled states need not satisfy these separability constraints, and infeasibility certifies entanglement (see~appendix  for more details).

%When  $\rho=\rho^{(m)}\otimes \rho^{(\bar m)}$ is  a product state, one has a positive semidefinite  $\Gamma=\Gamma^{(m)}\otimes\Gamma^{(\bar m)}$; when  state is  separable $\rho = \sum_h p_h \rho^{(m)}_h \otimes \rho^{(\bar{m})}_h$, $\Gamma$ is  a convex combination $\sum_h p_h \Gamma_h$ and positive semidefinite. For an entangled state, the matrix is not necessarily to be positive semidefinite (see SM).

%Using qubit measurement information, the non-physical terms can be expressed in terms of measurement parameters and experimental expectations. For instance, let $O^{(m)}_{e} =A_{0},  O^{(m)}_{e'}=A_{1}$ to be   qubit measurement, one has $A^{\dagger}_0 A_1 + A^{\dagger}_1 A_0 = 2\big( r'_1 A_0 + r'_0 A_1 + (r_0 r_1 c_{01} + r'_0 r'_1) \mathbb{I} \big)$, and its expectation values is expressed with measurement parameters and expectation values, unlike the general case where one has freedom to assign a real value to it.  If the state is separable, then some inferred measurement parameters and  physical terms can ensure that $\Gamma$ is positive semidefinite. Failure to do so implies entanglement.

In our setting, CBDI constrains certain non-physical entries through qubit operator identities. For example, for qubit observables one has
$A_0^\dagger A_1 + A_1^\dagger A_0 = 2\!\left(r'_1 A_0 + r'_0 A_1 + (r_0 r_1 c_{01}+r'_0 r'_1)\mathbb{I}\right)$, so its expectation value is fixed by the inferred measurement parameters and the observed correlators. Imposing these additional constraints reduces the freedom in $\Gamma$ and can make the separable feasibility problem infeasible, thereby certifying entanglement.

%\begin{equation}
%\label{eq:gamma_product_state} % 标签现在指向整个 equation 环境
%\begin{split}
%4\Gamma_{ee', ff'} &= {\rm Tr}\left[\rho^{(m)} \left(O^{(m)\dagger}_{e} {O}^{(m)}_{e'}+O^{(m)\dagger}_{e'} {O}^{(m)}_{e} \right)\right] \\
%&\quad \times {\rm Tr}\left[\rho^{\bar m}\left(O^{(\bar m)\dagger}_{f} {O}^{(\bar m)}_{f'}+O^{(\bar m)\dagger}_{f'} {O}^{(\bar m)}_{f}\right)\right] \\
%&= \Gamma^{(m)}_{ee'} \Gamma^{(\bar m)}_{ff'}, \quad \text{i.e., } \Gamma=\Gamma^{(m)}\otimes\Gamma^{(\bar m)}.
%\end{split}
%\end{equation} Substitute the expectation values
%As $\Gamma^{(m)}$ and $\Gamma^{(\bar m)}$ are positive, we have $\Gamma\ge0$.  
 %In the  {\emph{Example .3}?},   for any correlation  on the $x-y$ boundary  {(from our criterion?)}, the relevant matrix is  \emph{negative}  {nonpositive?}, which implies they cannot be realized by separable states in qubit system using independent devices. 
As an example, we  provide a  local correlation  in minimal Bell's scenario  defined with the expectation values     $\overline {A_{0}B_{0}}=\overline {A_{1}B_{1}}= 1 $,   $\overline{A_0B_1}=\overline {A_1B_0}=\cos \frac{\pi}{12}$,  and $\bar A_{0}=\bar A_{1}=\bar B_{0}=\bar B_{1}=0$, where $\overline {O}:= {\rm Tr}(\rho O)$.  
Assuming  qubit systems,  our criterion can uniquely determine these measurements  to be   PVMs  and $c_{01}=\cos \pi/12$ on both sides. 
Substituting these inferred parameters into the constrained moment-matrix test yields infeasibility under separability, thereby certifying entanglement.

\medskip
\noindent\textit{Conclusion and outlook.---}
We introduced a framework to analyze correlations produced by independent qubit devices in Bell and prepare-and-measure scenarios, where the absence of shared public randomness makes the attainable correlation sets typically non-convex. Using qubit-specific uncertainty relations, we derived a state-independent consistency criterion that captures such non-convexity, yields explicit boundaries in representative $2$-input/$2$-output examples, and enables correlation-based device inference from non-extreme data by constraining (and sometimes uniquely determining) unitary-invariant measurement parameters. We further showed that incorporating the inferred qubit measurement parameters  in a moment-matrix test can strengthen separability checks, allowing entanglement certification even for correlations that are Bell-local within the independent-device model.

Our approach complements established SDP/NPA-based linear tools for convex correlation sets and existing nonlinear, dimension-bounded witnesses based on determinants or purity. It directly exposes non-convex boundaries and enables parameter inference beyond the standard extreme-point  paradigm. Experimentally, these advances benefit current quantum technologies, where qubit devices are common and shared mixing resources are often absent. Our method offers efficient diagnostics of observed statistics, reduces ambiguity in device parameter inference, and  lower-overhead entanglement verification protocols that do not require operating at extreme or Bell-nonlocal points.

%Our  approach is a  complementary to well-established SDP/NPA-based linear tools that naturally address convex correlation sets and to existing nonlinear, dimension-bounded witnesses based on determinants or purity: it can expose non-convex boundaries directly and enable parameter inference beyond the usual extreme-point/self-testing paradigm.

%For extension to higher-dimensional systems, one may seek analogous criteria based on $d$-dimensional uncertainty relations that constrain unitary invariants of measurement families. For more parties, a key theoretical goal is to combine the conditional-state reduction with explicit global consistency conditions (beyond non-signaling) so that multipartite inference and boundary characterization become tighter and capture genuinely multipartite structure.

%On the experimental side, these developments could directly benefit current quantum technologies, where qubit devices are ubiquitous and shared mixing resources are often absent, by providing efficient diagnostics of observed statistics, reduced-ambiguity device parameter inference, and lower-overhead entanglement verification protocols that do not require operating at extreme or Bell-nonlocal points.

\textit{Acknowledgments}---
We appreciate the discussion with Armin Tavakoli. 
L.S.\ and S.Y.\ are supported by Quantum Science and Technology--National Science and Technology Major Project (2021ZD0300804). 
C.Z.\ is supported by the National Natural Science Foundation of China (Grant No.~62475127) and the Zhejiang Provincial Natural Science Foundation of China (Grant No.~LZ25A040006). 
Z.W.\ is supported by the NSFC (Grant No.~12574536) and the Sichuan Science and Technology Program (Grant No.~2024YFHZ0371).

\bibliography{pv}

\appendix
\onecolumngrid

% ============================================================
\section{Correlation criterion based on pairwise uncertainty relation}

Starting from the qubit uncertainty relation, we derive a correlation criterion that bounds the quantum correlations generated by \emph{independent} qubit systems.

\subsection*{PVM case}

Consider binary projection- valued  measures (PVMs)
\begin{equation}
  A_\ell=\vec a_\ell\cdot\vec\sigma, \qquad \lvert \vec a_\ell\rvert =1,
\end{equation}
and denote their expectation values on a state $\rho$ by
$A_{\ell|\rho}:=\operatorname{Tr}(\rho A_\ell)$ and $A^{2}_{\ell |\rho}=[{\rm Tr}(\rho A_{\ell})]^{2}$.
For any pair $(A_i,A_j)$ the qubit uncertainty relation reads
\begin{equation}
  A_{i|\rho}^{2}+A_{j|\rho}^{2}+c_{ij}^{2}-2c_{ij}A_{i|\rho}A_{j|\rho}\le 1,
\end{equation}
where $c_{ij}:=\vec a_j\cdot \vec a_i$.
Rearranging yields
\begin{equation}
  \big(c_{ij}-A_{i|\rho}A_{j|\rho}\big)^2\le (1-A_{i|\rho}^{2})(1-A_{j|\rho}^{2}),
\end{equation}
and hence
\begin{equation}
  g_-(A_{i|\rho},A_{j|\rho})\le c_{ij}\le g_+(A_{i|\rho},A_{j|\rho}),
  \label{qub1}
\end{equation}
with
\begin{equation}
  g_{\pm}(A_{i|\rho},A_{j|\rho})
  :=A_{i|\rho}A_{j|\rho}\pm \sqrt{1-A_{i|\rho}^{2}}\sqrt{1-A_{j|\rho}^{2}}.
\end{equation}

If the same measurement pair $(A_i,A_j)$ is tested on an ensemble of states
$\Omega=\{\rho_k\}$, then $c_{ij}$ is state-independent, while the interval in
Eq.~\eqref{qub1} depends on $\rho_k$ through the observed expectations.
A necessary condition for compatibility with a fixed $c_{ij}$ is therefore that the
intervals have a non-empty intersection:
\begin{equation}
\max_{\rho_k\in\Omega} g_-(A_{i|\rho_k},A_{j|\rho_k})
\ \le\
\min_{\rho_k\in\Omega} g_+(A_{i|\rho_k},A_{j|\rho_k}).
\label{eq:criterion_PVM}
\end{equation}

\subsection*{POVM case}

The same idea applies to general binary qubit measurements described by POVMs of the form
\begin{equation}
  A_\ell=r'_\ell\,\vec a_\ell\cdot\vec\sigma+r_\ell \openone,
  \qquad \lvert \vec a_\ell\rvert =1.
\end{equation}
Positivity of $(\openone\pm A_\ell)/2$ requires $\lvert r_\ell\rvert+r'_\ell\le 1$.
Without loss of generality we take $r'_\ell\ge 0$.
The normalized operator $(A_\ell-r_\ell\openone)/r'_\ell$ is then a PVM, so we can apply
Eq.~\eqref{qub1} to the rescaled expectations
$(A_{i|\rho}-r_i)/r'_i$ and $(A_{j|\rho}-r_j)/r'_j$.

Writing the upper bound explicitly gives
\begin{align}
c_{ij}
&\le \frac{A_{i|\rho}-r_i}{r'_i}\frac{A_{j|\rho}-r_j}{r'_j}
 +\sqrt{1-\Big(\frac{A_{i|\rho}-r_i}{r'_i}\Big)^2}
  \sqrt{1-\Big(\frac{A_{j|\rho}-r_j}{r'_j}\Big)^2}, \nonumber\\[2pt]
c_{ij}r'_ir'_j
&\le (A_{i|\rho}-r_i)(A_{j|\rho}-r_j)
 +\sqrt{{r'}_i^{\,2}-(A_{i|\rho}-r_i)^2}\,
  \sqrt{{r'}_j^{\,2}-(A_{j|\rho}-r_j)^2}.
\end{align}
Using $\bar r_\ell:=1-\lvert r_\ell\rvert\ge r'_\ell$ allows us to eliminate $r'_i$ and $r'_j$ from the
right-hand side:
\begin{equation}
c_{ij}r'_ir'_j
\ \le\
(A_{i|\rho}-r_i)(A_{j|\rho}-r_j)
+\sqrt{\bar r_i^{\,2}-(A_{i|\rho}-r_i)^2}\,
 \sqrt{\bar r_j^{\,2}-(A_{j|\rho}-r_j)^2}
=:g^{(\tilde\theta_4)}_{+}(A_{i|\rho},A_{j|\rho}),
\end{equation}
and analogously,
\begin{equation}
c_{ij}r'_ir'_j
\ \ge\
(A_{i|\rho}-r_i)(A_{j|\rho}-r_j)
-\sqrt{\bar r_i^{\,2}-(A_{i|\rho}-r_i)^2}\,
 \sqrt{\bar r_j^{\,2}-(A_{j|\rho}-r_j)^2}
=:g^{(\tilde\theta_4)}_{-}(A_{i|\rho},A_{j|\rho}).
\end{equation}
The constancy of $c_{ij}r'_ir'_j$ over $\rho_k\in\Omega$ thus leads to the POVM criterion
\begin{equation}
\max_{\rho_k\in\Omega} g^{(\tilde\theta_4)}_{-}\big(A_{i|\rho_k},A_{j|\rho_k}\big)
\ \le\
\min_{\rho_k\in\Omega} g^{(\tilde\theta_4)}_{+}\big(A_{i|\rho_k},A_{j|\rho_k}\big).
\label{g3a_3}
\end{equation}

% ------------------------------------------------------------
\subsection*{Example: Bell scenario}

We compare our uncertainty-relation (UR) criterion with representative criteria from the literature in the minimal
Bell scenario with two parties. Each party performs one of two binary measurements, $A_\alpha$ and $B_\beta$,
with outcomes $a,b\in\{0,1\}$ and inputs $\alpha,\beta\in\{0,1\}$.

We consider the correlation family
\begin{equation}
Q^{\mathrm{Bell}}_{xy}
= x\,\mathcal{P}_{\rm Pr} + y\,\mathcal{P}_{\rm L} + (1-x-y)\,\mathcal{P}_{\rm w},
\end{equation}
where
$\mathcal{P}_{\rm Pr}=\{p(ab|\alpha\beta)=\tfrac12,\ \forall\,a\oplus b=\alpha\beta\}$ is the PR box,
$\mathcal{P}_{\rm L}=\{p(00|\alpha\beta)=1,\ \forall\,\alpha,\beta\}$ is a local deterministic vertex,
and $\mathcal{P}_{\rm w}=\{p(ab|\alpha\beta)=\tfrac14,\ \forall\,a,b,\alpha,\beta\}$ is white noise.
Only a subset of $(x,y)$ can be generated by independent qubit systems; below we summarize how different criteria
constrain this region.

\paragraph*{Purity-based dimension witness.}

We now consider dimension witness of Ref.~\cite{PhysRevLett.117.060401}, which  is based on the fact that for a $d$-dimensional state $\rho_a$, its purity satisfies $\operatorname{Tr}(\rho_a^2) \geq 1/d$. Estimating the purity from experimental measurement statistics therefore yields a lower bound on the minimum Hilbert-space dimension necessary to describe the system.
\begin{align}
d\ge \max_{\beta,\beta'}\Bigg[\sum_{b,b'}\min_{\alpha}
\left(\sum_{a}\sqrt{p(ab|A_{\alpha}B_{\beta})}\sqrt{p(ab'|A_{\alpha}B_{\beta'})}\right)^{2}\Bigg]^{-1}=: w(x,y).
\end{align}
Evaluating $w(x,y)$ on $Q^{\mathrm{Bell}}_{xy}$ gives the corresponding $x$--$y$ boundary in the main text by imposing
$w(x,y)\le 2$. (The closed-form expression for $w(x,y)$ is lengthy and therefore omitted.)

\paragraph*{NPA hierarchy.}
The NPA hierarchy~\cite{PhysRevLett.98.010401} provides SDP outer approximations to the quantum set
(without assuming a fixed Hilbert-space dimension). In the minimal Bell scenario, one convenient form is
\begin{equation}
\left|\sum_{\alpha,\beta}(-1)^{\alpha\beta}\arcsin D_{\alpha\beta}\right|\le \pi,
\end{equation}
where
\begin{equation}
D_{\alpha\beta}
=\frac{\overline{A_{\alpha}B_{\beta}}-\overline{A_{\alpha}}\ \overline{B_{\beta}}}
{\sqrt{(1-\overline{A_{\alpha}}^{2})(1-\overline{B_{\beta}}^{2})}}.
\label{eq:D_alpha_beta_def}
\end{equation}
Here the correlators and marginals are
\begin{align}
\overline{A_{\alpha}B_{\beta}}
:=\sum_{a,b}(-1)^{a+b}\,p(ab|A_{\alpha}B_{\beta}),\qquad
\overline{A_{\alpha}}:=\sum_{a,b}(-1)^{a}\,p(ab|A_{\alpha}B_{\beta}),\qquad
\overline{B_{\beta}}
:=\sum_{a,b}(-1)^{b}\,p(ab|A_{\alpha}B_{\beta}).
\end{align}
For the family $Q^{\mathrm{Bell}}_{xy}$ this reduces to
\begin{equation}
\left|\,3\arcsin\frac{y+x-x^2}{1-y^2}-\arcsin\frac{y-x-x^2}{1-y^2}\,\right|\le \pi.
\end{equation}

% ------------------------------------------------------------
\subsection*{Example: prepare-and-measure scenario}

We also consider a prepare-and-measure (PM) correlation $Q^{\mathrm{PM}}_{xy}$ obtained as a modification of
$Q^{\mathrm{Bell}}_{xy}$. It is constructed from the observable set $\mathcal{O}=\{A_0,A_1\}$ and the state ensemble
$\Omega=\{\rho_{b|\beta}\}$, where $\rho_{b|\beta}$ denotes the (subnormalized) state prepared for Alice conditioned on Bob's input $\beta$ and outcome $b$.

A $d$-dimensional quantum state is characterized by $d^2-1$ parameters; therefore, any set of $d^2$ or more states in dimension $d$ must be linearly dependent. This motivates a nonlinear dimension witness based on $2k$ states and $k$ binary measurements~\cite{PhysRevLett.112.140407}. Defining a $k \times k$ matrix $\mathcal{W}_k$ with entries $w_{mn}=p(0|m,\rho_{2n})-p(0|m,\rho_{2n+1})$, where $m$ denotes the measurement setting, its determinant satisfies $|\det \mathcal{W}_k| = 0$ when $d \le \sqrt{k}$, while $|\det \mathcal{W}_k| > 0$ becomes possible for $d > \sqrt{k}$.

Labeling Bob's events as $b|\beta\in\{0|0,1|0,0|1,1|1\}$, the conditional expectation values on Alice's side are $\{\rho_{b|\beta}\}$ we have  
\begin{equation}
A_{\alpha|\rho_{b|\beta}}
=\frac{p(0b|\alpha\beta)-p(1b|\alpha\beta)}{p(b|\beta)},
\qquad
p(b|\beta)=\sum_{a}p(ab|\alpha\beta),
\end{equation}
where the sum runs over Alice's outcomes (and $p(b|\beta)$ does not depend on $\alpha$ by no-signalling).  
For $Q^{\mathrm{PM}}_{xy}$ the relevant expectation-value pairs can be written as
\begin{align*}
    & \left\{ \frac{1+x+3y}{2+2y},\ \frac{1+x+3y}{2+2y} \right\},\quad
      \left\{ \frac{1-x-y}{2-2y},\ \frac{1-x-y}{2-2y} \right\}, \\
    & \left\{ \frac{1+x+3y}{2+2y},\ \frac{1-x+3y}{2+2y} \right\},\quad
      \left\{ \frac{1-x-y}{2-2y},\ \frac{1+x-y}{2-2y} \right\}.
\end{align*}

For the determinant-based criterion with $k=2$,
\begin{equation}
\mathcal{W}_2=
\begin{pmatrix}
p(0|0,\rho_{0|0})-p(0|0,\rho_{1|0}) & p(0|0,\rho_{0|1})-p(0|0,\rho_{1|1}) \\[4pt]
p(0|1,\rho_{0|0})-p(0|1,\rho_{1|0}) & p(0|1,\rho_{0|1})-p(0|1,\rho_{1|1})
\end{pmatrix},
\end{equation}
where  $p(0|\alpha, \rho_{b|\beta})={(1+A_{\alpha|\rho_{b|\beta}})}/{2}$. For $d=2$ the condition $\lvert\det\mathcal{W}_2\rvert\le 1$ yields the boundary
\begin{equation}
\bigl|2xy-2xy^{2}+2x^{2}\bigr|\le (1-y^{2})^{2}.
\end{equation}

% ============================================================
\section{CBDI protocol details}

In this section we summarize the construction of the correlation-based device inference (CBDI) protocol and provide
explicit inference steps, together with conditions under which the relevant device parameters are uniquely determined.

\begin{proposition}[Polynomial-time uniqueness detection for a measurement pair]
\label{prop:poly_unique}
Fix a pair of binary measurements $(A_i,A_j)$ on a qubit and a finite set $\Omega=\{\rho_k\}_{k=1}^N$.
Assume the observed expectation values $A_{i|\rho_k}$ and $A_{j|\rho_k}$ are given as rational numbers
(or as reals specified to $L$ bits of precision).
Define the compatible parameter set $\mathcal{S}$ of unitary-invariant POVM parameters
$\theta_{5}=(r_i,r'_i,r_j,r'_j,c_{ij})$ by the qubit uncertainty-relation constraints (given below).
Then one can decide in time $\mathrm{poly}(N,L)$ whether $\mathcal{S}\neq\emptyset$
(membership) and whether $|\mathcal{S}|=1$ (unique determination).
For PVMs, the same holds with $c_{ij}\in[-1,1]$ and the runtime is $O(N)$.
\end{proposition}

\begin{proof}[Explicit algorithm]
We give an algorithm whose runtime is polynomial in the number of states $N$ and the input precision $L$.
The key point is that for qubits the number of unknown unitary-invariant parameters per measurement pair is a fixed constant.

\paragraph*{Step 0 (Input).}
For each $\rho_k\in\Omega$, record the pair $(A_{i|\rho_k},A_{j|\rho_k})$.

\paragraph*{Step 1 (PVM case: interval intersection, $O(N)$).}
If $A_i,A_j$ are binary PVMs, they can be written as $A_\ell=\vec a_\ell\cdot\vec\sigma$, and the only
unitary-invariant parameter of the pair is the overlap $c_{ij}=\vec a_i\cdot\vec a_j\in[-1,1]$.
For each $k$, the qubit uncertainty relation implies
\begin{equation}
c_{ij}\in\bigl[g_-(A_{i|\rho_k},A_{j|\rho_k}),\ g_+(A_{i|\rho_k},A_{j|\rho_k})\bigr].
\end{equation}
Hence $\mathcal{S}\neq\emptyset$ if and only if
\begin{equation}
\max_{\rho_k \in \Omega} g_-\ \le\ \min_{\rho_k\in \Omega} g_+,
\end{equation}
where we have suppressed the arguments of $g_{\pm}$ for brevity.
Moreover, $|\mathcal{S}|=1$ if and only if the inequality is saturated
(up to a chosen numerical tolerance). Computing the two extrema is $O(N)$.

\paragraph*{Step 2 (POVM case: semi-algebraic feasibility in constant dimension).}
For general binary qubit POVMs, write
\begin{align}
A_\ell&=r'_\ell\,\vec a_\ell\cdot\vec\sigma+r_\ell\,\openone, \quad
-1\le r_\ell\le 1,\qquad 0\le r'_\ell\le 1-\lvert r_\ell\rvert,\qquad \ell\in\{i,j\}.
\end{align}
Let $c_{ij}:=\vec a_i\cdot\vec a_j\in[-1,1]$, and define for each $k$
\begin{equation}
\tilde A_{i|\rho_k}:=A_{i|\rho_k}-r_i,\qquad \tilde A_{j|\rho_k}:=A_{j|\rho_k}-r_j,
\end{equation}
i.e., $\tilde A_\ell=(A_\ell-r_\ell\openone)$.
Applying the PVM uncertainty relation to the normalized observables $\tilde A_\ell/r'_\ell$ yields
\begin{equation}
\bigl(c_{ij}r'_ir'_j-\tilde A_{i|\rho_k}\tilde A_{j|\rho_k}\bigr)^2
\le \bigl(r_i^{\prime 2}-\tilde A_{i|\rho_k}^2\bigr)\bigl(r_j^{\prime 2}-\tilde A_{j|\rho_k}^2\bigr),
\label{eq:S1}
\end{equation}
together with domain constraints ensuring the square roots are well-defined:
\begin{equation}
r_i^{\prime 2}-\tilde A_{i|\rho_k}^2\ge 0,\qquad
r_j^{\prime 2}-\tilde A_{j|\rho_k}^2\ge 0,\qquad k=1,\dots,N.
\label{eq:S2}
\end{equation}
Finally, the parameter ranges are
\begin{align}
-1\le r_i,r_j\le 1,
\quad 0\le r'_i\le 1-\lvert r_i\rvert, \quad 0\le r'_j\le 1-\lvert r_j\rvert, \quad-1 \le c_{ij}\le 1.
\label{eq:S3}
\end{align}
Equations \eqref{eq:S1}--\eqref{eq:S3} define a semi-algebraic set $\mathcal{S}\subset\mathbb{R}^5$ described by
$O(N)$ polynomial inequalities of constant degree (at most $4$), up to the absolute values in \eqref{eq:S3}.

\paragraph*{Step 3 (Eliminate absolute values by constant branching).}
To eliminate $\lvert r_i\rvert$ and $\lvert r_j\rvert$, branch over the four sign choices
$(s_i,s_j)\in\{(+1,+1),(+1,-1),(-1,+1),(-1,-1)\}$ and enforce $s_ir_i\ge 0$ and $s_jr_j\ge 0$.
In each branch, replace $\lvert r_i\rvert$ by $s_ir_i$ and $\lvert r_j\rvert$ by $s_jr_j$.
This produces four purely polynomial systems, each still in $5$ variables and with $O(N)$ constraints.

\paragraph*{Step 4 (Membership test).}
For each branch, decide whether the corresponding polynomial inequality system has a real solution.
This is an instance of deciding emptiness of a semi-algebraic set in \emph{fixed dimension} (here $n=5$ variables),
with $s=O(N)$ constraints of constant degree and coefficients specified with $L$ bits.
Standard algorithms for semi-algebraic feasibility in fixed dimension (e.g., quantifier elimination for the existential
theory of the reals in fixed dimension) decide this in time $\mathrm{poly}(s,L)=\mathrm{poly}(N,L)$.
If any branch is feasible, then $\mathcal{S}\neq\emptyset$.

\paragraph*{Step 5 (Unique determination detection).}
Unique determination (for the invariant parameters) means that $\mathcal{S}$ contains exactly one point.
Equivalently, $\mathcal{S}$ is not a singleton if and only if there exist two distinct solutions
$\theta_{5},\theta'_{5}\in\mathcal{S}$ with $\theta_{5}\neq\theta'_{5}$.
This can be checked by solving a second fixed-dimensional feasibility problem: introduce two copies of variables
\[
\theta_{5}=(r_i,r'_i,r_j,r'_j,c_{ij}),\qquad
\theta'_{5}=(\hat r_i,\hat r'_i,\hat r_j,\hat r'_j,\hat c_{ij}),
\]
impose the constraints \eqref{eq:S1}--\eqref{eq:S3} on both copies (with the same experimental data), and add the strict
inequality
\begin{equation}
\|\theta_{5}-\theta'_{5}\|_2^2
=(r_i-\hat r_i)^2+(r'_i-\hat r'_i)^2+(r_j-\hat r_j)^2+(r'_j-\hat r'_j)^2+(c_{ij}-\hat c_{ij})^2>0.
\label{eq:S4}
\end{equation}
As in Step~3, remove absolute values by branching over the signs of $(r_i,r_j,\hat r_i,\hat r_j)$, yielding a constant
number of branches ($2^4=16$). Each branch is a polynomial feasibility instance in $10$ variables (still constant
dimension) with $O(N)$ constraints and $L$-bit coefficients. Therefore, each branch can be decided in time
$\mathrm{poly}(N,L)$, and so can the overall uniqueness test.

\paragraph*{Step 6 (From invariant uniqueness to measurement uniqueness up to unitary).}
For binary qubit measurements, the tuple $\theta_{5}$ fixes the pair $(A_i,A_j)$ up to the natural gauge freedoms:
a global unitary acting on the qubit and outcome relabeling ($A\mapsto -A$ together with swapping outcomes).
Thus, $|\mathcal{S}|=1$ provides an algorithmic certificate of ``unique determination up to unitary.''

\paragraph*{Runtime.}
The PVM routine is $O(N)$.
For POVMs, the algorithm performs a constant number of fixed-dimensional semi-algebraic feasibility checks, each with
$O(N)$ constraints and $L$-bit coefficients, hence runs in $\mathrm{poly}(N,L)$ time.
\end{proof}

\subsection*{Example: Device inference}

We consider correlations on the boundary of the $x$--$y$ plane for $Q^{\mathrm{PM}}_{xy}$.
On this boundary one can (for the family studied in the main text) uniquely determine the measurements
\begin{equation}
A_0=\sigma_z,\qquad
A_1=r_1c_{01}\sigma_z+r_1\sqrt{1-c_{01}^2}\,\sigma_y+(1-r_1)\openone,
\end{equation}
up to an overall unitary, where analytic expressions for $c_{01}(x)$ and $r_1(x)$ are given below.
Note that these boundary correlations are typically non-extremal (except at the endpoints), yet they can still uniquely
fix the measurement pair.

We now turn to the Bell-family correlations $Q^{\mathrm{Bell}}_{xy}$ and study the boundary induced by qubit states
together with general two-outcome qubit measurements. We assume
\begin{equation}
A_\alpha=r_\alpha\,\vec a_\alpha\cdot\vec\sigma+r'_\alpha\,\openone
=r_\alpha A_\alpha^{(0)}+r'_\alpha\openone,
\qquad \lvert r_\alpha\rvert+\lvert r'_\alpha\rvert\le 1,
\end{equation}
and use the shorthand $A_{\alpha|\rho_{b|\beta}}=:A_{\alpha|\beta_b}$.

\begin{itemize}
\item \emph{Boundary ansatz.}
Numerical evidence suggests that on the relevant boundary one can take
\begin{equation}
r_0=1,\quad r'_0=0,\qquad r_1=r,\quad r'_1=1-r.
\end{equation}
We adopt this as an assumption for the analytic derivation below. Under this ansatz the conditional averages for the
ideal observables (PVM parts) are
\begin{equation}
A^{(0)}_{\alpha|\beta_b}
=\frac{1}{1-\alpha r'_\alpha}\left(\frac{x\bigl(1+(-1)^b\bigr)+t(-1)^{b+\alpha\beta}}{1+(-1)^b x}-\alpha r'_\alpha\right).
\end{equation}

\item \emph{Boundary condition.}
The boundary is determined by the equalities
$g_{-,0_0}=g_{-,0_1}=g_{+,1_1}$, where
\begin{align}
g_{-,\beta_b}
&:=A_{0|\beta_b}A_{1|\beta_b}-\sqrt{(1-A_{0|\beta_b}^2)(1-A_{1|\beta_b}^2)},\nonumber\\
g_{+,\beta_b}
&:=A_{0|\beta_b}A_{1|\beta_b}+\sqrt{(1-A_{0|\beta_b}^2)(1-A_{1|\beta_b}^2)}.
\end{align}

\item \emph{Parametrization by the measurement angle.}
Let $c_{01}:=\vec a_0\cdot\vec a_1\equiv\cos\theta$. Solving
$g_{-,0_0}=g_{+,0_1}=g_{+,1_1}=\cos\theta$ yields
\begin{equation}
\frac1x
=1+\frac{1}{\cos\theta}\frac{2(1+\sin\theta)(1-\cos\theta)}{2+\cos\theta+\sqrt{(1+\sin\theta)(1+3\sin\theta)}},
\label{xthe}
\end{equation}
and
\begin{equation}
x=(1-y)\frac{\sqrt{1+3\sin\theta}-\sqrt{1-\sin\theta}}{2\sqrt{1+\sin\theta}}.
\label{xythe}
\end{equation}

\item \emph{State reconstruction.}
For a qubit state $\rho=\tfrac12(\openone+\vec s\cdot\vec\sigma)$ and two PVMs
$A_0=\vec a_0\cdot\vec\sigma$ and $A_1=\vec a_1\cdot\vec\sigma$ with expectations $A_0$ and $A_1$,
the Bloch vector can be written as
\begin{equation}
\vec s=
\frac{1}{\sin^2\theta}\,(A_0,A_1)
\begin{pmatrix}
1&-\cos\theta\\
-\cos\theta&1
\end{pmatrix}
\begin{pmatrix}
\vec a_0\\
\vec a_1
\end{pmatrix}
+t\,\frac{\vec a_0\times\vec a_1}{\lvert\vec a_0\times\vec a_1\rvert^{2}}.
\label{lit}
\end{equation}

\item \emph{Purity constraints and the free parameter $t$.}
For $\beta_b\in\{0_0,0_1,1_1\}$ the boundary states satisfy $\lvert\vec s_{\beta_b}\rvert=1$  and $t=0$, i.e.,
\begin{equation}
\lvert\vec s_{\beta_b}\rvert^2
=\frac{1}{\sin^2\theta}\,(A^{(0)}_{0|\beta_b},A^{(0)}_{1|\beta_b})
\begin{pmatrix}
1&-\cos\theta\\
-\cos\theta&1
\end{pmatrix}
\begin{pmatrix}
A^{(0)}_{0|\beta_b}\\
A^{(0)}_{1|\beta_b}
\end{pmatrix}
=1.
\label{eq:purity_constraint_boundary}
\end{equation}
In the remaining case $\beta_b=1_0$, the state $\rho_{0|1}$ can have an arbitrary $t$ subject to
\begin{equation}
|t|\le \sqrt{(1-A_{0|1_0}^2)(1-A_{1|0_1}^2)-(\cos\theta-A_{0|0_1}A_{1|0_1})^2}.
\label{eq:t_condition}
\end{equation}
\end{itemize}

Finally, the parameter $r$ can be expressed in terms of
\begin{align}
r = \frac{u}{\sqrt{1-u^2}} - \frac{u^2}{1-u^2}, 
\quad u = \frac{x}{1-y} = \frac{\sqrt{1+3\sin\theta} - \sqrt{1-\sin\theta}}{2\sqrt{1+\sin\theta}}.
\end{align}

\begin{figure}
\includegraphics[scale=0.45]{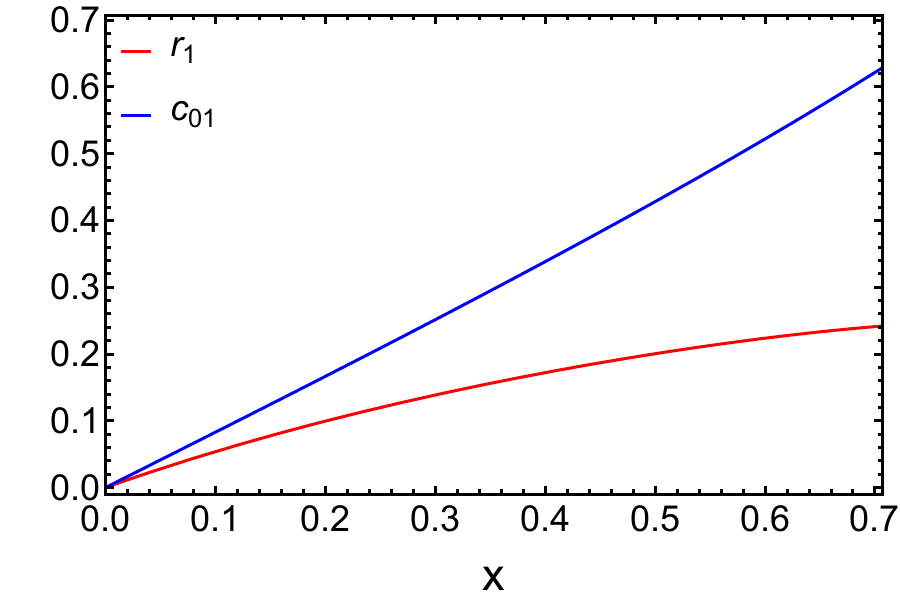}
\caption{Plots of $r_{1}$ and $c_{01}$. Along the boundary, the measurements of $A_0$ and $A_1$ are uniquely determined
up to a global unitary as $A_0=\sigma_z$ and
$A_1=r_1c_{01}\sigma_z+r_1\sqrt{1-c_{01}^2}\,\sigma_y+(1-r_1)\openone$.}
\end{figure}

\subsection*{Example: simple protocols}

The above analysis provides evidence that CBDI can uniquely determine devices using simpler experimental setups than
purely linear approaches. In particular, the minimal protocol discussed in the main text employs four states and three
measurements.

Consider correlations with expectation-value pairs
\begin{align*}
     \left\{ \frac{1+x+3y}{2+2y},\ \frac{1+x+3y}{2+2y} \right\},\quad
      \left\{ \frac{1-x-y}{2-2y},\ \frac{1-x-y}{2-2y} \right\}, \quad \left\{ \frac{1-x-y}{2-2y},\ \frac{1+x-y}{2-2y} \right\},
\end{align*}
where $x$ and $y$ are related by Eqs.~\eqref{xthe} and \eqref{xythe}. From these correlations one can determine the
state ensemble and measurement set described in the previous subsection, namely
$\mathcal{O}=\{A_0,A_1\}$ and $\Omega=\{\rho_{0|0},\rho_{1|0},\rho_{1|1}\}$.

A simpler protocol involves two states and two measurements:
\begin{equation}
A_{0|\rho_0}=A_{1|\rho_0}=1,\qquad A_{0|\rho_1}=-A_{1|\rho_1}=1.
\end{equation}
Our criterion is satisfied only when $A_0=\openone$ and $A_1$ is a projective measurement, with the two states being the
eigenstates of $A_1$. Although the identity operator $\openone$ is mathematically a valid measurement operator (it can be
interpreted as verifying whether a particle has been recorded), it is physically trivial. It remains an open question
whether two states and two measurements can be used to certify \emph{non-trivial} qubit measurements.

% ============================================================
\section{Optimization of Entanglement-detection in qubit systems}

The NPA hierarchy defines a sequence of tests for deciding whether an observed correlation can be produced by a quantum
state and measurements. Fix an operator list $\mathcal{O}=\{O_e\}$, where each $O_e$ is a measurement operator or a product thereof. For any linear combination $W=\sum_e c_e O_e$ with coefficient vector
$\mathbf{C}=(c_1,\dots,c_{n'})^T$, one has
\begin{equation}
\operatorname{Tr}(\rho W^\dagger W)=\mathbf{C}^\dagger \Gamma\,\mathbf{C}\ \ge\ 0,
\end{equation}
where the moment matrix $\Gamma$ has entries
\begin{equation}
\Gamma_{ef}:=\operatorname{Tr}\!\left(\rho\,O_e^\dagger O_f\right).
\end{equation}
Therefore $\Gamma$ must be positive semidefinite.

Some entries of $\Gamma$ correspond to directly measurable statistics. For example, if $O_e=A_e$ and $O_f=B_f$ are
measurement operators of two parties, then $\Gamma_{ef}=\operatorname{Tr}(\rho A_eB_f)$ is an observable correlator.
Other entries, such as $\operatorname{Tr}(\rho A_e^\dagger A_f)$ for noncommuting operators $[A_e,A_f]\neq 0$, are not
associated with any directly measurable quantity; we refer to them as \emph{non-physical} terms. If a correlation is
quantum-realizable, one can assign values to these non-physical terms so that the resulting $\Gamma$ is positive semidefinite. Conversely, if no such assignment exists, then the correlation is not quantum-realizable.

To test separability one may exploit additional structure. Consider a bipartition $(m|\bar m)$ and take an operator list
of the form $O_{ef}=O^{(m)}_e\otimes O^{(\bar m)}_f$. Define a symmetrized moment matrix via
\begin{align}
4\Gamma_{ee',ff'}
=\operatorname{Tr}\Big[\rho\big(O^{(m)\dagger}_e O^{(m)}_{e'}+O^{(m)\dagger}_{e'}O^{(m)}_e\big)\times\big(O^{(\bar m)\dagger}_f O^{(\bar m)}_{f'}+O^{(\bar m)\dagger}_{f'}O^{(\bar m)}_f\big)\Big].
\label{eq:gamma_def_split}
\end{align}
For a product state $\rho=\rho^{(m)}\otimes \rho^{(\bar m)}$ this factorizes as
\begin{align}
4\Gamma_{ee',ff'}
&=\operatorname{Tr}\!\left[\rho^{(m)}\big(O^{(m)\dagger}_e O^{(m)}_{e'}+O^{(m)\dagger}_{e'}O^{(m)}_e\big)\right]
\times
\operatorname{Tr}\!\left[\rho^{(\bar m)}\big(O^{(\bar m)\dagger}_f O^{(\bar m)}_{f'}+O^{(\bar m)\dagger}_{f'}O^{(\bar m)}_f\big)\right]=\Gamma^{(m)}_{ee'}\,\Gamma^{(\bar m)}_{ff'},
\end{align}
i.e., $\Gamma=\Gamma^{(m)}\otimes \Gamma^{(\bar m)}$.
Since $\Gamma^{(m)}\succeq 0$ and $\Gamma^{(\bar m)}\succeq 0$, their tensor product is also positive semidefinite.
For a general separable state $\rho=\sum_h p_h\,\rho^{(m)}_h\otimes\rho^{(\bar m)}_h$, the corresponding moment matrix is a
convex combination $\sum_h p_h \Gamma_h\succeq 0$. This need not hold for entangled states.

For qubit systems one can further strengthen the test by incorporating the inferred measurement information from
Appendix~B. For instance, at the $1.5$ level of the NPA hierarchy, one may take
\begin{equation}
\tilde{\mathcal{O}}=\{\openone,\ A^{(m)}_{\alpha},\ B^{(\bar m)}_{\beta},\ A^{(m)}_{\alpha}B^{(\bar m)}_{\beta}\}.
\end{equation}
When $O^{(m)}_e=A_0$ and $O^{(m)}_{e'}=A_1$ are qubit observables, the anticommutator can be expressed as
\begin{equation}
A_0^\dagger A_1 + A_1^\dagger A_0
=2\Big(r'_1 A_0 + r'_0 A_1 + \bigl(r_0r_1c_{01}+r'_0r'_1\bigr)\openone\Big),
\end{equation}
whose expectation value depends only on operational terms and constants already certified by the CBDI inference
constraints. Substituting admissible parameters $\theta_5\in\mathcal{S}$ into $\Gamma$ therefore fixes entries that are
otherwise treated as free non-physical variables. Entanglement is certified if $\Gamma$ fails to be positive semidefinite
for \emph{all} $\theta_5\in\mathcal{S}$ consistent with the observed data.

\end{document}